\documentclass{PoS}

\usepackage{graphicx}
\usepackage{amsmath,amssymb}
\usepackage{capt-of}

\newcommand{\ba}{\begin{eqnarray}}
\newcommand{\ea}{\end{eqnarray}}
\newcommand{\be} {\begin{equation}}
\newcommand{\ee} {\end{equation}}

\title{Calculation of $K\to\pi l \nu$ 
form factors with $N_f=2+1$ flavours  of staggered quarks}

\ShortTitle{$K\to\pi l \nu$ form factors with staggered quarks}

\author{
\speaker{E.~G\'amiz}$^{a,b}$,
C.~DeTar$^c$,
A.X.~El-Khadra$^d$,
A.S.~Kronfeld$^b$,
P.B.~Mackenzie$^b$, 
and 
J.~Simone$^b$\\ \\
\llap{$^a$}CAFPE and Depto. de F\'{\i}sica Te\'orica y del Cosmos, 
Universidad de Granada, E-18002, Granada, Spain \\
\llap{$^b$}Fermi National Accelerator Laboratory,\hspace*{-0.4em}
    \thanks{Operated by Fermi Research Alliance, LLC, under Contract
    No.~DE-AC02-07CH11359 with the United States Department of Energy.}~
 Batavia, IL 60510, USA \\
\llap{$^c$}Department of Physics and Astronomy, University of Utah, Salt Lake City, UT
84112, USA \\
\llap{$^d$}Physics Department, University of Illinois, Urbana, IL  61801,
USA \\

E-mail: \email{megamiz@ugr.es}}
\author{Fermilab Lattice and MILC Collaborations\\
}

\abstract{
We report on the status of the Fermilab-MILC calculation of the form factor 
$f_+^{K \pi}(q^2=0)$, needed to extract the CKM matrix element $|V_{us}|$ from 
experimental data on $K$ semileptonic decays. The HISQ formulation is used 
in the simulations for the valence quarks, while the sea quarks are simulated 
with the asqtad action (MILC $N_f=2+1$ configurations). We discuss the general 
methodology of the calculation, including the use of twisted boundary conditions 
to get values of the momentum transfer close to zero and the different techniques 
applied for the correlators fits. We present initial results for lattice spacings 
$a\approx 0.12~{\rm fm}$ and $a\approx 0.09~{\rm fm}$, and several choices of the light 
quark masses.
}

\FullConference{ The XXIX International Symposium on Lattice Field Theory - 
Lattice 2011\\
July 10-16, 2011\\
Squaw Valley, Lake Tahoe, California}

\begin{document}

\section{Introduction}

The error associated with the lattice determination of the form factor 
$f_+^{K\pi}(0)$ ($\sim 0.5\%$) is still the dominant uncertainty in the 
extraction of $\vert V_{us}\vert$ from experimental data on $K$ semileptonic 
decays:  $\vert V_{us}\vert f_+^{K\pi}(0) = 0.2163(\pm 0.23\%)$ 
\cite{Antonellietal2010}. 
Improvement in the determination of that form factor is thus crucial  
in order to extract all the information from the available experimental data. 

A precise value of $\vert V_{us}\vert$ is needed to check unitarity in the 
first row of the CKM matrix. Any deviation from unitarity would indicate 
the existence of beyond the Standard Model physics. But, even if unitarity is 
fulfilled, however, as it is the case with current experimental and theoretical inputs, 
this test can establish very stringent constraints on the scale 
of the allowed new physics ($\sim 10~{\rm TeV}$) \cite{Antonellietal2010}. 
One also could compare the values of $\vert V_{us}\vert$ as extracted from 
helicity-allowed semileptonic decays and helicity-suppressed leptonic decays  
in the search for deviations from SM predictions. In particular, it is 
useful to study the ratio 
\ba \label{eq:Rmu23}
R_{\mu 23} = \left(\frac{f_+^{K\pi}(0)}{f_K/f_\pi}\right)_{{\rm lattice}}
\left(\left|\frac{V_{us}}{V_{ud}}\right|\frac{f_K}{f_\pi}\right)_{\mu 2}\,
\frac{\vert V_{ud}\vert}{\left\lbrack\vert V_{us}\vert 
f_+^{K\pi}(0)\right\rbrack_{l3}}\, ,
\ea
where the subscripts $\mu 2$ and $l 3$ indicate that those quantities are 
obtained from experimental data on leptonic $K_{\mu 2}$ and semileptonic $K_{l 3}$ 
decays respectively. The ratio in (\ref{eq:Rmu23}) is unity in the SM but not in some 
extensions of the SM, for example, those 
with a charged Higgs. Again, the error in the current value $R_{\mu 23} = 0.999(7)$ 
\cite{Antonellietal2010} is limited by the precision of lattice-QCD inputs.

In these proceedings we report on the status of the calculation of the form factor 
$f_+^{K\pi}(0)$ using staggered quarks. The goal of this analysis is to show that 
the staggered formulation can provide a determination of this parameter competitive 
with the state of the art unquenched determinations \cite{latticef+} by addressing 
the main sources of systematic errors and improving in statistics.

\section{Methodology: extracting the form factor directly at $q^2=0$}

Semileptonic $K$ decays are parametrized in terms of the form factors $f_+$ 
and $f_0$ in the following way
\ba\label{eq:formfac}
\langle \pi \vert V^\mu \vert K\rangle = 
f_+^{K\pi}(q^2) \left[p_K^\mu + p_\pi^\mu - \frac{m_K^2-m_\pi^2}{q^2}          
q^\mu\right]+f_0^{K\pi}(q^2)\frac{m_K^2-m_\pi^2}{q^2}q^\mu\,,
\ea
where $q=p_K-p_\pi$ and $V^\mu$ is the appropriate flavour changing vector current. 
One of the main components of our analysis which reduces both systematic and 
statistical errors is the use of the method developed by the HPQCD 
collaboration to study charm semileptonic decays \cite{HPQCD_Dtopi}. 
This method is based on the Ward identity relating the matrix 
element of a vector current to that of the corresponding scalar current:  
$q^\mu\langle \pi\vert V_\mu^{lat.} \vert K\rangle Z=(m_s-m_q)
\langle \pi\vert S^{lat.} \vert K\rangle\,$,
with $S=\bar s l$, and $Z$ a lattice renormalization factor for the vector
current. In this work, we use the local scalar density  of staggered fermions, so 
the combination $(m_s-m_q)S$ requires no renormalization. 
Using the definition of the form factors in Eq.~(\ref{eq:formfac})
and this identity, one can extract
$f_0^{K\pi}(q^2)$ at any $q^2$
by using
\ba\label{eq:WIresult}
f_0^{K\pi}(q^2) = \frac{m_s-m_l}{m_K^2-m_\pi^2}\langle \pi\vert S
\vert K\rangle (q^2).
\ea
Kinematic constraints demand that $f_+(0)=f_0(0)$, so this relation can be
used to calculate $f_+^{K\pi}(0)$. One of the main advantages of relation 
(\ref{eq:WIresult}) is that it avoids the use of a renormalization factor to 
obtain the form factor $f_0$. The drawback to this method is that it
gives no access to the shape of $f_+^{K\pi}$, but in this analysis we are focusing 
on the extraction of $\vert V_{us}\vert$, so we need the normalization of 
the form factor only at a single point. 

\begin{center}
\begin{figure}[tb]
\begin{minipage}[c]{0.48\textwidth}
\begin{center}

\includegraphics[width=0.95\textwidth]{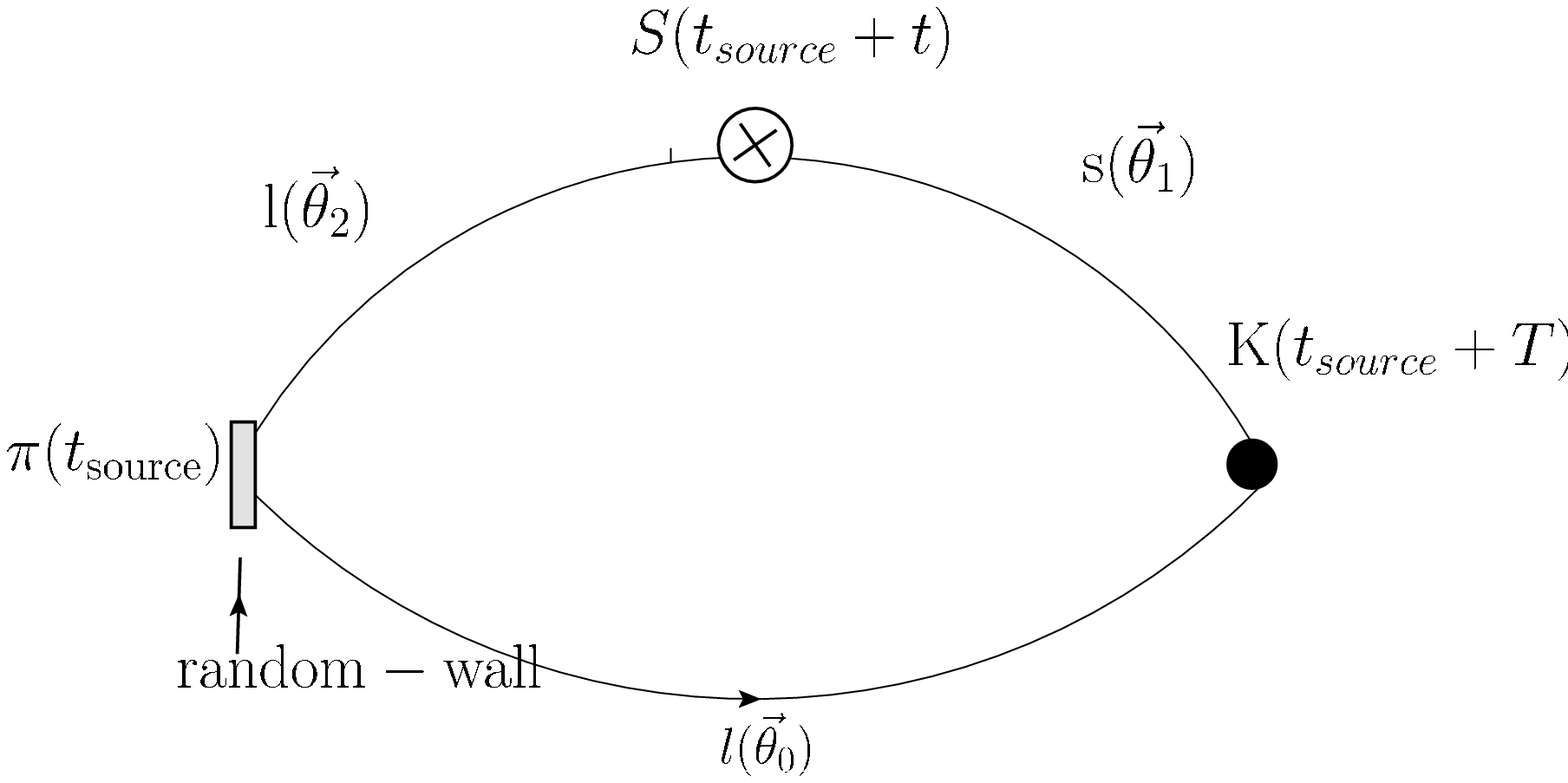}
\end{center}
\captionof{figure}{Structure of the 3-point functions needed to
calculate $f_{0}^{K\pi}(q^2)$. Light-quark propagators are
generated at $t_{source}$ with random-wall sources. An 
extended strange propagator is generated at $T$. 
\label{fig:diagram}}
\end{minipage}
\hspace*{0.4cm}
\begin{minipage}[c]{0.48\textwidth}
\begin{center}
\vspace*{-0.5cm}
\includegraphics[angle=-90,width=0.85\textwidth]{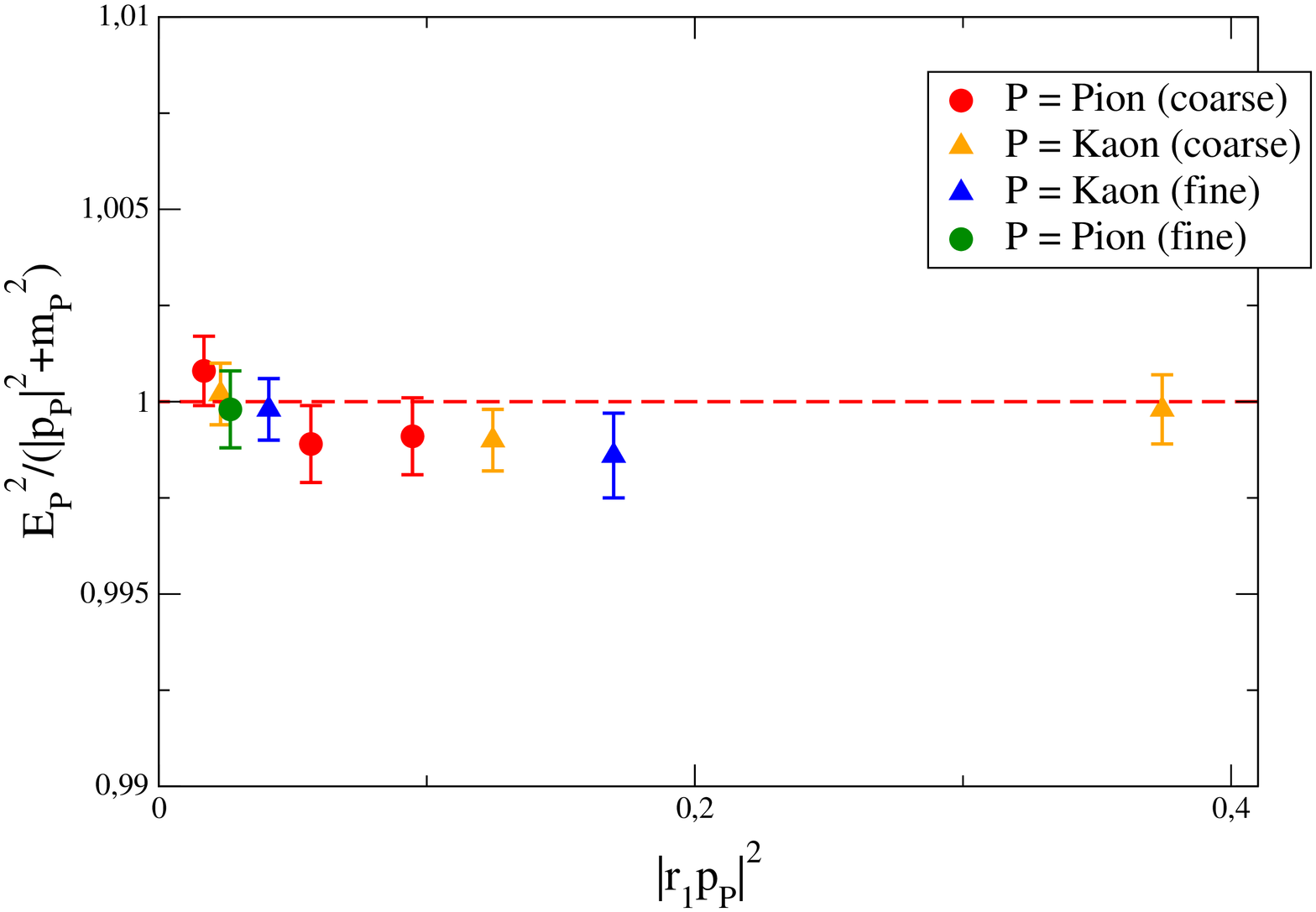}
\captionof{figure}{Deviation of our data from the continuum dispersion relation 
prediction. \label{fig:dispersion}}
\end{center}
\end{minipage}
\end{figure}
\end{center}

Another key ingredient is employing twisted boundary 
conditions~\cite{Bedaque:2004ax,Boyle:2010bh} 
to simulate the relevant correlations functions directly at $q^2=0$. This avoids   
an interpolation in $q^2$ and thus the corresponding systematic 
uncertainty. In Fig.~\ref{fig:diagram} we plot the general structure of the 
relevant 3-point functions. In order to get $q^2=0$, 
we inject momentum $\vec{p}=\vec{\theta}\pi/L$ in either the kaon or the pion. 
For a non-zero $\vec{p}_K$ we chose $\vec{\theta}_0=\vec{\theta}_2=0$,
$\vec{\theta}_1\ne\vec{0}$, and for a non-zero $\vec{p}_\pi$ we chose $\vec{\theta}_0=
\vec{\theta}_1=\vec{0}$, $\vec{\theta}_2\ne\vec{0}$ (see Fig.~\ref{fig:diagram} 
for definition of $\vec{\theta}_{0,1,2}$). The twisting angles  
are tuned to produce $q^2=0$ using two-point correlators fits 
according to 
\ba\label{twistangle}
\vec{\theta}_1 (q^2=0) =  \frac{L}{\pi}\, \sqrt{\left(
\frac{m_K^2+m_\pi^2}{2m_\pi}\right)^2-m_K^2}\,\, ,\quad\quad
\vec{\theta}_2 (q^2=0) =  \frac{L}{\pi}\, \sqrt{\left(
\frac{m_K^2+m_\pi^2}{2m_K}\right)^2-m_\pi^2}\, .
\ea

\vspace*{-0.2cm}
We found in a previous test run \cite{latt2010} that the use of random-wall sources 
greatly reduces the statistical errors of the parameters of the two-point and 
three-point correlators, so we use them throughout this analysis.

\section{Simulation details and fitting}

We have completed the generation of correlators with HISQ staggered valence quarks on the 
$2+1$ flavor asqtad staggered MILC ensembles~\cite{MILCasqtad} shown in 
Table~\ref{tab:sim} at two lattice spacings.
\begin{table}[tb]
\begin{center}
\begin{tabular}{lccccccc}
\hline\hline
       & $\approx a$ (fm) & $am_l/am_h$ & $N_{conf}$
& $am_s^{val}$ & $am_l^{val}$ & $N_{sources}$ & $N_T$ \\
\hline
coarse & $0.12$ & $0.020/0.050$ & $2052$ & $0.0491(9)$ & $0.02806$ & $4$ & $5$ \\
       &        & $0.010/0.050$ & $2243$ & $0.0495(9)$ & $0.01414$ & $4$ & $8$ \\
       &        & $0.005/0.050$ & $2098$ & $0.0489(9)$ & $0.00670$ & $8$ & $5$ \\
\hline
fine   & $0.09$ & $0.0124/0.031$ & $1996$ & $0.0337(6)$ & $0.0080$ & $4$ & $5$ \\

        &       & $0.0062/0.031$ & $1946$ & $0.0336(6)$ & $0.0160$ & $4$ & $5$ \\
\hline\hline
\end{tabular}
\end{center}
\caption{Ensembles and simulation details. $am_h$ is the nominal strange-quark mass 
in the sea sector, $N_{sources}$ is the number of time sources, and $N_T$ the 
number of sink-source separations for which we have generated data. \label{tab:sim}}
\end{table}
We average results over four time sources separated by 16 (24) timeslices on the 
$0.12~{\rm fm}$ ($0.09~{\rm fm}$) ensembles but displaced by a random distance 
from configuration to configuration to suppress autocorrelations. A subset of this 
data was analyzed in~\cite{latt2010}. The strange valence mass is tuned to 
its physical value on each ensemble~\cite{r1HPQCD}. 
The valence light-quark masses are fixed according to the relation, 
$\frac{m_l^{val}(HISQ)}{m_s^{phys.}(HISQ)} =\frac{m_l^{sea}(asqtad)}{m_s^{phys.}
(asqtad)}$. 
The effect of the mixed actions for the sea and valence quark sectors can be 
analyzed using partially quenched staggered CHPT techniques for the chiral and 
continuum extrapolations.

We fit the two-point functions for a pseudoscalar meson $P$ to the expression
\ba\label{2pt}
C_{2pt}^{P} (\vec{p}_P;t) & = & \sum_{m=0}^{N_{exp}} (-1)^{mt}(Z_m^P)^2
\left(e^{-E_P^m t}+e^{-E_P^m(L_t-t)}\right)\,,
\ea
where $L_t$ is the temporal size of the lattice. Oscillating terms with 
$(-1)^m$ do not appear for pions with zero momentum. From two-point function  
fits, we checked whether the continuum dispersion relation is satisfied. 
This is plotted in Fig.~\ref{fig:dispersion}, which shows very small deviations 
from the continuum prediction ($\le 0.15\%$), indicating small discretization 
effects.

The functional form for the three-point functions is
\ba\label{3pt}
        C_{3pt}^{K\to \pi} (\vec{p}_\pi,\vec{p}_K;t,T) & = &
\sum_{m,n=0}^{N_{exp}^{3pt}} (-1)^{mt} (-1)^{n(T-t)}A^{mn}Z_m^\pi Z_n^K
\left(e^{-E_\pi^mt-E_\pi^m(L_t-t)}\right)
\left(e^{-E_K^n(T-t)-E_K^n(T-L_t+t)}\right)\,,\nonumber\\
\ea

\vspace*{-0.3cm}
\noindent where the factors $Z_i^P$ are the amplitudes of the two-point functions in (\ref{2pt}).
The three-point parameter $A^{00}$ in (\ref{3pt}) is related to 
the desired form factor $f_0^{K\pi}(q^2)$ via  
$f_0^{K\pi}(q^2) = \frac{1}{2}A^{00}(q^2)\, \sqrt{2E_\pi E_K}\,
(m_s-m_l)/(m_K^2-m_\pi^2)$, where we have used (\ref{eq:WIresult}) and taken into 
account some overall factors involved in the parametrization of the correlation 
function. We extract the form factors $f_0^{K\pi}(q^2)$ using the expression above 
directly from simultaneous fits of the relevant three- and two-point functions. 
In these fits we include several three-point functions with different values 
of the source-sink separation $T$, with at least one odd $T$ and one even $T$ to be 
able to get a handle on the contributions from the oscillatory states.

In this analysis it is especially relevant to check for the stability of our fits 
under the choice of fitting parameters and techniques, since we are getting very small 
statistical errors and we need to be sure that these results are not methodology 
dependent in any way. One of the checks we performed is varying the time fitting 
ranges and number of states included in the fits. Fitting ranges for 
two-point functions are $t_{min}-(L_t-t_{min})$ and for the three-point functions
$t_{min}-(T-t_{min})$, with $L_t$ the temporal size of the lattice and $T$ the
source-sink separation---see Fig.~\ref{fig:diagram}. The number of states 
included is the same in the regular and oscillating sectors, so
$N_{exp}=N_{{\rm regular\,states}}=N_{{\rm oscillatory\,states}}$. Fixing $N_{exp}$ 
and changing $t_{min}$ from 3 (5) for $0.12~{\rm fm}$ ($0.09~{\rm fm}$) 
ensembles up to the maximum allowed by the source-sink separation, give us a 
plateau for central  values with only small variations in errors. Analogously, 
fixing $t_{min}$ to our preferred value we do not find any significant variation 
of results for $N_{exp}\ge 3-4$. 

We study as well which combination of $T's$ from the ones we have simulated is optimal.
We find that the central values are very insensitive to the number of three-point functions
included and the values of $T$ in the range we are analyzing. Errors and stability are
better when $ 15\le T \le 24$ and including three three-point functions for 
the $0.12~{\rm fm}$ ensembles, and $ 18\le T \le 33$ and including four three-point 
functions for the $0.09~{\rm fm}$ ensembles. 

Finally, we checked an alternative way of doing the fits, using the iterative
superaverage method described in \cite{Btopi}. This takes an explicit
combination of three-point functions with consecutive values of $T$ and the time 
slice $t$ which suppresses the contribution from both the first regular 
excited state and the first oscillatory state. Again, results are compatible 
within one statistical $\sigma$ with our preferred fitting method.

\begin{center}
\begin{figure}[tbh]
\begin{minipage}[c]{0.48\textwidth}
\begin{center}
\vspace*{-0.5cm}
\includegraphics[angle=-90,width=1.\textwidth]{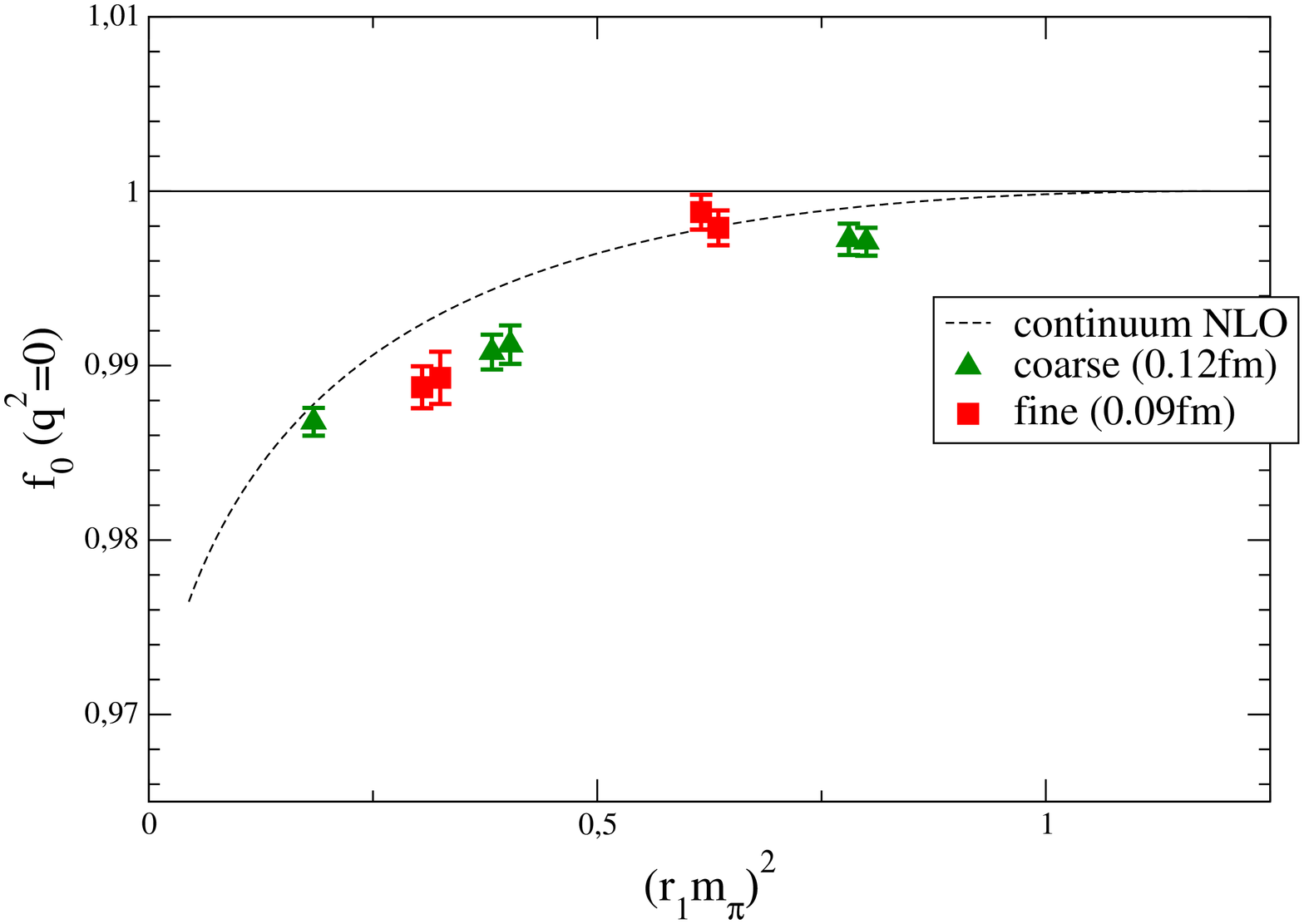}
\end{center}
\end{minipage}
\hspace*{0.5cm}
\begin{minipage}[c]{0.48\textwidth}
\begin{center}
\vspace*{-0.5cm}
\includegraphics[angle=-90,width=1.\textwidth]{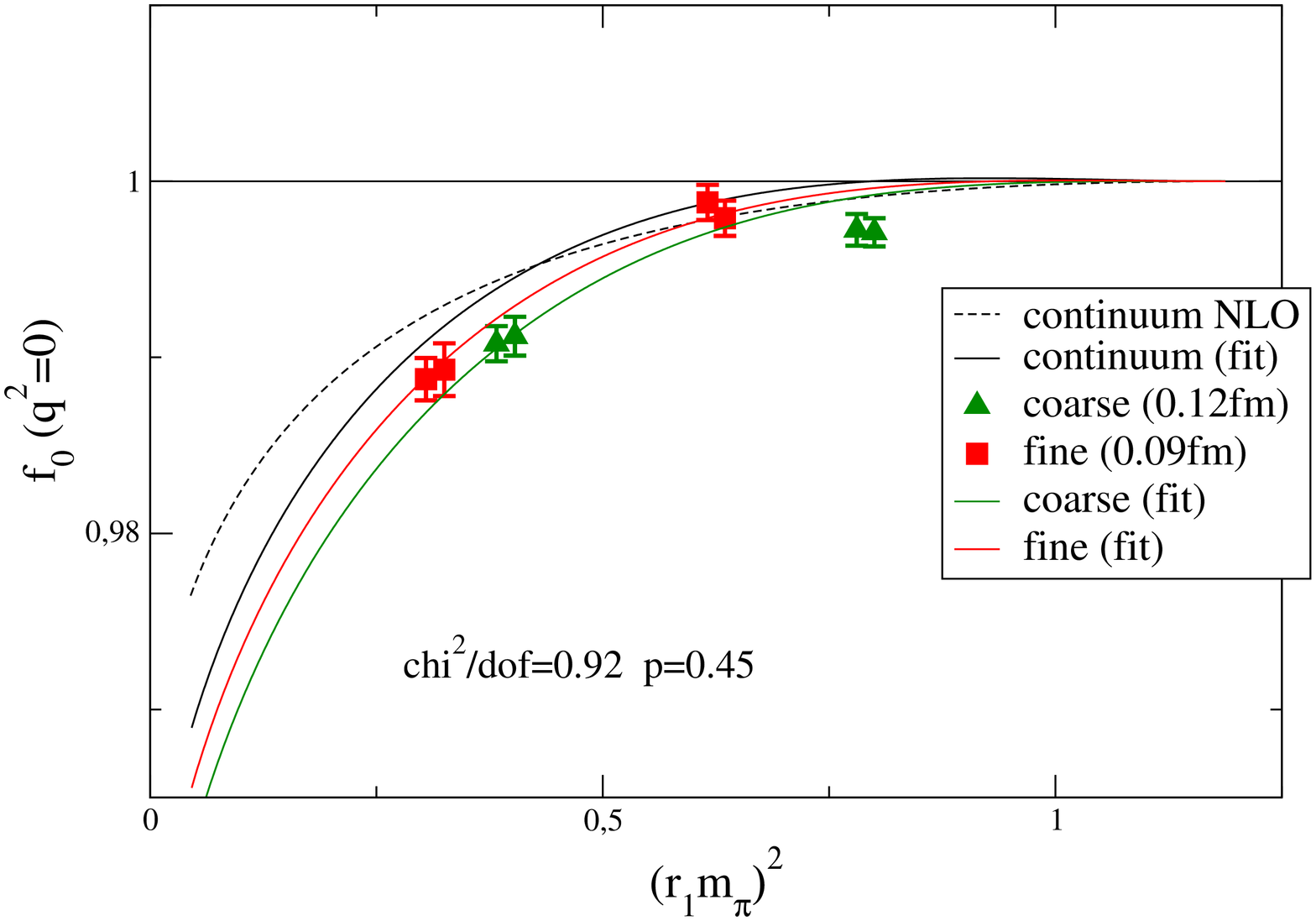}
\end{center}
\end{minipage}
\caption{In the left panel, we collect the form factor $f_+^{K\pi}(0)$ obtained
from the different ensembles in Table~[1]. The right panel shows an example of the
output from the chiral and continuum extrapolation tests. The dotted black line in
both plots is the NLO CHPT prediction. 
The solid black line is the extrapolation in the light-quark mass with a physical $m_s$
and in the continuum limit. The green and red lines are also extrapolations in the 
light-quark masses with $m_s^{phys}$ but with $a\approx 0.12~{\rm fm},0.09~{\rm fm}$, 
respectively.
\label{fig:results}}
\end{figure}
\end{center}
\vspace*{-0.4cm}

\section{Results for $f_+(q^2=0)$}

In Fig.~\ref{fig:results} we collect the results for $f_+^{K\pi}(0)$ with statistical 
errors from our preferred fits as a function of $(r_1m_\pi)^2$, fixing 
$(2m_K^2-m_\pi^2)$ to the experimental value. We plot results coming from three-point 
functions where the external momentum to obtain $q^2=0$ is injected via 
the $K$ and the $\pi$ (we give an offset in the pion mass to the two points for 
clarity). The green triangle at the far left corresponds  to the $0.12~{\rm fm}$ 
ensemble with masses 0.005/0.050, four time sources, and a moving pion. 
When analyzing those data we found it to be more challenging to get stable results, 
so we have decided to double the number of sources and exclude it in our discussions 
until the full new data set is analyzed and other effects, like finite volume 
corrections, are incorporated to the analysis. So, in particular, we do not include 
it in the fits described below.

The first remarkable characteristic of our results is that the statistical errors are
very small, $0.1-0.15\%$, reaching our goal to be competitive with other 
determinations. In addition, 
the results coming from three-point functions where the external momentum to 
get $q^2=0$ is injected via the $K$ and the $\pi$ agree within the very small 
statistical errors, as can be seen in the 
figure. This constitutes a very good test of our methodology and quoted errors.

\subsection{Chiral and continuum extrapolation}

The form factor $f_+(0)$ can be written as a CHPT expansion in the following way: 
$f_+(0) = 1 + f_2 + f_4 + f_6 + ...= 1 + f_2 + \Delta f$. The Ademollo-Gatto (AG) 
theorem, which follows from vector current conservation, ensures that 
$f_+(0) \to 1$ in the $SU(3)$ limit and, furthermore, that the $SU(3)$  
breaking effects are second order in $(m_K^2-m_\pi^2)$. This fixes $f_2$ completely 
in terms of experimental quantities. At finite lattice spacing, systematic  errors 
can enter due to, for example, corrections to the dispersion relation needed to derive 
Eq.~(\ref{eq:WIresult}). 
Those and other discretization effects are very small though, as can be 
deduced from our results in Fig.~\ref{fig:results}. However, since 
statistical errors are at the $0.1-0.2\%$ level, we should pin down the other 
sources of systematic errors as precisely as possible. 

Our plan for treating the light-quark mass dependence and the discretization effects 
in our calculation is to use two-loop continuum CHPT \cite{BT03}, 
supplemented by staggered partially quenched CHPT at one-loop. 
The small variation with $a$ in our data suggests that addressing those effects at 
one loop should be enough for our target precision. 

Since we do not yet have the staggered CHPT expressions, nor have we implemented 
the two-loop continuum CHPT functions, just as an exercise, we can try to fit our 
data with a much more simple fitting ansatz. We take the continuum partially quenched 
NLO CHPT expression \cite{Bijnenspc} and add a general parametrization of 
NNLO analytic terms and $a^2$ corrections of the form 
\ba \label{eq:ChPTfunction}
f_+^{K\pi}(0) = 1 + f_2 + C_a \left(\frac{a}{r_1}\right)^2 + &&r_1^4(m_\pi^2-m_K^2)^2
\Big\lbrack C_6^{(1)}(r_1 m_\pi)^2 + C_6^{(2)}(r_1 m_K)^2\nonumber\\
+ && C_6^{(3)}(r_1 m_\pi)^2log(m_\pi^2/\mu^2) + C_6^{(4)} (r_1 m_\pi)^4 +
C_6^{(5)}\left(\frac{a}{r_1}\right)^2\Big\rbrack.
\ea
We include only correlation functions coming from injecting
the momentum in the $\pi$ for these test fits. The result for one of these fits 
with $C_a=C_6^{(i)}=0$ for $i=3,4$ is shown in Fig.~\ref{fig:results}. We can obtain  
similar good fits with different combinations of terms in (\ref{eq:ChPTfunction}).   
This must be taken just as a naive first try to fit our data and no conclusions 
should be drawn until we have used staggered CHPT to gain more information about 
the $a^2$ structure of our data.

\section{Conclusions and outlook}

We have completed the generation of the data in Table~\ref{tab:sim} needed for the 
calculation of $f_+^{K\pi}(0)$. Since the time of the conference we have generated 
data for another coarse ensemble with light-quark mass $am_l=0.007$ to facilitate  
the chiral extrapolation, which we anticipate is going to be our main source of 
uncertainty.

The statistical errors in the form factor in all ensembles exceed our expectations, 
being around $0.1-0.15\%$. We have performed several checks of the robustness of the 
central fit values and errors, by studying the stability with changes in the time 
range and number of states, the dependence on the source-sink separation and number 
of three-point functions included in the fit, and testing alternative methods for 
fitting the correlation functions.  
We find it very difficult to make changes in the fitting procedure that change the 
fit results outside the one sigma range. Another very good test of our results 
is the fact that $f_+^{K\pi}(0)$ as extracted from three-point correlation functions with 
a moving $\pi$ and a moving $K$ agree with each other. For the final analysis we will 
redo the fits we found to be the optimal, including the correlation functions with 
both momentum injected in the $\pi$ and the $K$ to further increase the statistics.

We found very small lattice spacing dependence in our data and the continuum
dispersion relation is fulfilled at the $0.15\%$ level, but in view of the small
statistical error, we plan to study in detail the dependence on $a^2$ by using staggered
partially quenched CHPT at one-loop. We will also investigate the use of two-loop 
continuum CHPT.

With all these elements, we expect our calculation to be competitive with the  
current state of the art. 

\begin{acknowledgments}
Computations for this work were carried out with resources provided by the 
USQCD Collaboration, the Argonne Leadership Computing
Facility, the National Energy Research Scientific Computing Center, and 
the Los Alamos National Laboratory, which are funded by the Office of Science of 
the U.S. Department of Energy; and with resources provided by the National 
Institute for Computational Science, the Pittsburgh Supercomputer Center, the San Diego 
Supercomputer Center, and the Texas Advanced Computing Center, which are funded 
through the National Science Foundation's Teragrid/XSEDE Program. 
This work was supported in part by the MICINN (Spain) under grant FPA2010-16696 
and \emph{Ram\'on y Cajal} program (E.G.), Junta de Andaluc\'{\i}a (Spain) under 
grants FQM-101, FQM-330, FQM-03048, and FQM-6552 (E.G.), 
and by the U.S. Department of Energy under Grant No. DE-FG02-91ER40677 (A.X.E.).

\end{acknowledgments}

\end{document}